\documentclass[letterpaper]{jpconf}
\usepackage{ae}
\usepackage{bm}
\usepackage{color}
\usepackage{amsmath}
\usepackage{amssymb}
\usepackage{amsfonts}
\usepackage{graphicx}
\usepackage{slashed}
\usepackage{wasysym}
\usepackage{setspace}
\usepackage{cite}

\begin{document}

\title{Photon propagation in slowly varying electromagnetic fields}

\author{Felix Karbstein}

\address{Helmholtz-Institut Jena
\& Theoretisch-Physikalisches Institut, Abbe Center of Photonics, \\ Friedrich-Schiller-Universit\"at Jena, Max-Wien-Platz 1, 07743 Jena, Germany}

\ead{felix.karbstein@uni-jena.de}

\begin{abstract}
 We study the effective theory of soft photons in slowly varying electromagnetic background fields at one-loop order in QED.
 This is of relevance for the study of all-optical signatures of quantum vacuum nonlinearity in realistic electromagnetic background fields as provided by high-intensity lasers.
 The central result derived in this article is a new analytical expression for the photon polarization tensor in two linearly polarized counter-propagating pulsed Gaussian laser beams.
 As we treat the peak field strengths of both laser beams as free parameters this field configuration can be considered as interpolating between the limiting cases of a purely right- or left-moving laser beam
 (if one of the peak field strengths is set to zero) and the standing-wave type scenario with two counter-propagating beams of equal strength.
\end{abstract}

\section{Introduction}

Apart from the classic result for infinitely extended constant electromagnetic fields \cite{Heisenberg:1935qt}, only a few exact results for the one-loop effective action in specific (one-dimensional) field inhomogeneities are known explicitly, cf., e.g., \cite{Cangemi:1995ee,Dunne:1997kw,Dunne:1998ni}, and \cite{Dunne:2004nc} for a review. The effective action vanishes identically for the case of a single monochromatic plane wave field \cite{Schwinger:1951nm}.
For the one-loop photon polarization tensor, analytical expressions are only known for constant \cite{BatShab} and plane-wave fields \cite{Baier:1975ff,Becker:1974en}.
Contrarily, the field configurations delivered by high-intensity lasers do not belong to any of these classes \cite{DiPiazza:2011tq}. Typically, they are rather spatially and temporally confined and inhomogeneous \cite{Siegman}. 

\section{The locally constant field approximation}

Many experimentally relevant field configurations are amenable to a locally constant field approximation (LCFA):
In the limit of low-frequency fields, the typical distanced probed by virtual electron-positron fluctuations are of the order of the electron Compton wavelength $\lambda_c=3.86\cdot10^{-13}{\rm m}$.
By definition the scales of variation $\lambda$ of low-frequency fields are large in comparison to $\lambda_c$.
Hence, these fields do not change much over distances of the order of $\lambda_c$ and can be approximated as locally constant.
In constant electromagnetic fields the entire field dependence of the effective action $S\equiv S({\cal F},{\cal G}^2)$ is in terms of the gauge and Lorentz invariants of the electromagnetic field: ${\cal F}=\frac{1}{4}F_{\mu\nu}F^{\mu\nu}=\frac{1}{2}(\vec{B}^2-\vec{E}^2)$ 
and ${\cal G}=\frac{1}{4}F_{\mu\nu}^*F^{\mu\nu}=-\vec{E}\cdot\vec{B}$. Here $^*F^{\mu\nu}=\frac{1}{2}\epsilon^{\mu\nu\alpha\beta}F_{\alpha\beta}$ denotes the dual field strength tensor; $\epsilon^{\mu\nu\alpha\beta}$ is the totally antisymmetric tensor, fulfilling $\epsilon^{0123}=1$. As the effective action is a scalar quantity and $\cal G$ is a pseudoscalar, the latter dependence is in terms of ${\cal G}^2$.
Our metric convention is $g_{\mu \nu}=\mathrm{diag}(-1,+1,+1,+1)$, and we use units where $c=\hbar=1$.
The LCFA amounts to substituting $F^{\mu\nu}\to F^{\mu\nu}(x)$, such that the effective action becomes a functional of $F^{\mu\nu}(x)$,
\begin{equation}
 S({\cal F},{\cal G}^2)=\int{\rm d}^4x\, {\cal L}({\cal F},{\cal G}^2)\quad \to\quad S[{\cal F}(x),{\cal G}^2(x)]=\int{\rm d}^4x\, {\cal L}\bigl({\cal F}(x),{\cal G}^2(x)\bigr)\,. \label{eq:StoSx}
\end{equation}
Of course, the LCFA inherently neglects derivative terms, and the resulting effective Lagrangian ${\cal L}\bigl({\cal F}(x),{\cal G}^2(x)\bigr)$ is fully local.
Nevertheless, this approximation grasps the essential features of the field inhomogeneity.
Deviations from the unknown full result are expected to be suppressed parametrically by powers of the ratio $\lambda/\lambda_c$ (cf. \cite{Karbstein:2015cpa} for more details).

\section{The effective theory for photon propagation}

The effective action can be written as $S[{\cal F}(x),{\cal G}^2(x)]=-\int{\rm d}^4x\, {\cal F}(x)+S_{\rm qc}[{\cal F}(x),{\cal G}^2(x)]$,
where the first term is the Maxwell action of classical electrodynamics.
The additional term $S_{\rm qc}=\int{\rm d}^4x\, {\cal L}_{\rm qc}$ encodes quantum corrections and vanishes in the limit $\hbar\to0$.
Decomposing the field strength tensor as $F^{\mu\nu}(x)\to F^{\mu\nu}(x) + f^{\mu\nu}(x)$ into a slowly varying background field $F^{\mu\nu}(x)$ and a dynamical soft photon field
$f^{\mu\nu}(x)= {\rm i}\int\frac{{\rm d}^4k}{(2\pi)^4}\,{\rm e}^{{\rm i}kx}\bigl[k^\mu g^{\nu\sigma} - k^\nu g^{\mu\sigma}\bigr]a_\sigma(k)$,
the quantity ${\cal L}_{\rm qc}$ can serve as generator of the effective soft photon interactions mediated by quantum fluctuations of virtual electrons and positrons.
Here, we exclusively focus on the effective action at one-loop order, such that ${\cal L}_{\rm qc}$ corresponds to the one-loop Heisenberg-Euler effective Lagrangian \cite{Heisenberg:1935qt} (for reviews cf. \cite{Dittrich:1985yb}). In this limit, the effective $n$-photon interaction arises from the following interaction term in the effective action,
\begin{equation}
 S^{(n)}_{\rm int}=\frac{1}{n!}\int{\rm d}^4x\,\biggl(\prod_{j=1}^nf^{\mu_j\nu_j}(x)\frac{\partial}{\partial F^{\mu_j\nu_j}(x)}\biggr){\cal L}_{\rm qc}\bigl({\cal F}(x),{\cal G}^2(x)\bigr)\,.
\end{equation}
At linear and quadratic order in $a_\sigma(k)$ this generically gives rise to a photon current,
\begin{equation}
 j^\sigma(k) \equiv\frac{\delta S_{\rm qc}}{\delta a_\sigma(k)}\biggl|_{a=0} \,, \label{eq:j0}
\end{equation}
induced by the background field,
and to the photon polarization tensor,
\begin{equation}
 \Pi^{\rho\sigma}(k,k') \equiv -\frac{\delta^2 S_{\rm qc}}{\delta a_\rho(k)a_\sigma(k')}\biggl|_{a=0} \,, \label{eq:Pi0}
\end{equation}
mediating between two distinct photon momenta $k^\mu$ and $k'^\mu$.

The expressions for the induced photon current and the photon polarization tensor in a background field configuration characterized by the single field strength tensor $F^{\mu\nu}(x)$ \cite{Karbstein:2015cpa} can be straightforwardly generalized
to a field configuration that can be decomposed as $F^{\mu\nu}(x)=\sum_{i=1}^N F^{\mu\nu}_i(x)$, where $F^{\mu\nu}_i(x)$ denote the individual field strength tensors of $N\geq1$ superposed background fields.
Writing the background field invariants in terms of the individual field strength tensors $F^{\mu\nu}_i(x)$, we obtain
\begin{equation}
 {\cal F}(x)=\frac{1}{4}\sum_{i,j=1}^N F_{i\,\mu\nu}(x) F_j^{\mu\nu}(x)\quad \text{and}\quad {\cal G}(x)=\frac{1}{4}\sum_{i,j=1}^N F_{i\,\mu\nu}(x) {}^*F_j^{\mu\nu}(x) \,,
\end{equation}
where both indices $i$ and $j$ are summed over all integers from $1$ to $N$. 
In order to keep notations compact, we henceforth omit the explicit reference to $x$ for the field strength tensors and the argument of the effective Lagrangian.
Moreover, we make use of the shorthand notation $(kF_i)^\mu=k_\nu F^{\nu\mu}_i$, $(k{}^*F_i)^\mu=k_\nu{}^*F^{\nu\mu}_i$ and $(kk')=k_\mu k'^\mu$.
In turn, the induced photon current and the photon polarization tensor can be expressed as
\begin{equation}
 j^\rho(k) = {\rm i}\int{\rm d}^4x\,{\rm e}^{{\rm i}kx} \sum_{i=1}^N\biggl[ (kF_i)^\rho\,\frac{\partial{\cal L}_{\rm qc}}{\partial{\cal F}}
 + (k{}^*F_i)^\rho\,\frac{\partial{\cal L}_{\rm qc}}{\partial{\cal G}}\biggr]\,, \label{eq:j}
\end{equation}
and
\begin{align}
 \Pi^{\rho\sigma}(k,k')
 =\int{\rm d}^4x\,{\rm e}^{{\rm i}(k+k')x}&\biggl\{
 \bigl((k k')g^{\rho\sigma} - k'^\rho k^\sigma \bigr)\frac{\partial{\cal L}_{\rm qc}}{\partial{\cal F}}
 + \epsilon^{\rho\sigma\alpha\beta} k'_\alpha k_\beta\, \frac{\partial{\cal L}_{\rm qc}}{\partial{\cal G}} \nonumber\\
 &\ + \sum_{i,j=1}^N\biggl[(kF_i)^\rho  (k'F_j)^\sigma\,\frac{\partial^2 {\cal L}_{\rm qc}}{\partial{\cal F}^2}
 + (k{}^*F_i)^\rho (k'{}^*F_j)^\sigma\,\frac{\partial^2 {\cal L}_{\rm qc}}{\partial{\cal G}^2} \nonumber\\
 &\ + \bigl[(k{}^*F_i)^\rho (k'F_j)^\sigma + (k F_i)^\rho (k'{}^*F_j)^\sigma\bigr]\,\frac{\partial^2 {\cal L}_{\rm qc}}{\partial{\cal F}\partial{\cal G}}\biggr]
 \biggr\}\,. \label{eq:Pi}
\end{align}

Essentially all electromagnetic fields available in the laboratory fulfill $\frac{e\epsilon}{m^2}\ll1$, where we count ${\cal O}(F^{\mu\nu}_i)\sim{\cal O}(\epsilon)$.
This suggests a perturbative expansion of Eqs.~\eqref{eq:j} and \eqref{eq:Pi} in powers of the background field strength $\epsilon$.
In particular note that
\begin{equation}
 \left\{\begin{array}{c} \frac{\partial{\cal L}_{\rm qc}}{\partial{\cal F}}\vspace*{1mm} \\ \frac{\partial{\cal L}_{\rm qc}}{\partial{\cal G}} \end{array}\right\}  = \frac{\alpha}{\pi}\frac{1}{45}\Bigl(\frac{e}{m^2}\Bigr)^2
 \left\{\begin{array}{c} 4{\cal F} \\ 7{\cal G} \end{array}\right\}+{\cal O}\bigl((\tfrac{e\epsilon}{m^2})^4\bigr)\,.
\end{equation}
Correspondingly, the leading contribution to the photon current~\eqref{eq:j} is cubic, and the photon polarization tensor~\eqref{eq:Pi} is quadratic in the background field strength. We obtain
\begin{equation}
 j^\rho(k) = {\rm i}\frac{\alpha}{\pi}\frac{1}{45}\Bigl(\frac{e}{m^2}\Bigr)^2
 \int{\rm d}^4x\,{\rm e}^{{\rm i}kx} \sum_{i=1}^N\biggl[ (kF_i)^\rho\,4{\cal F} 
 + (k{}^*F_i)^\rho\,7{\cal G}\biggr] + {\cal O}(\epsilon^5)\,, \label{eq:j1}
\end{equation}
and
\begin{multline}
 \Pi^{\rho\sigma}(k,k')
 =\frac{\alpha}{\pi}\frac{1}{45}\Bigl(\frac{e}{m^2}\Bigr)^2\int{\rm d}^4x\,{\rm e}^{{\rm i}(k+k')x}\biggl\{
 4{\cal F}\bigl((k k')g^{\rho\sigma} - k'^\rho k^\sigma \bigr)
 + 7{\cal G}\epsilon^{\rho\sigma\alpha\beta} k'_\alpha k_\beta \\
 + \sum_{i,j=1}^N\Bigl[4(kF_i)^\rho  (k'F_j)^\sigma
 + 7(k{}^*F_i)^\rho (k'{}^*F_j)^\sigma 
 \Bigr]
 \biggr\} + {\cal O}(\epsilon^4)\,. \label{eq:Pi1}
\end{multline}

\section{The photon polarization tensor in two counter-propagating Gaussian beams}

Here, we aim at providing an explicit expression of the photon polarization tensor in the presence of $N=2$ background fields $F_\pm^{\mu\nu}(x)$,
which we assume to be provided by two counter-propagating pulsed Gaussian laser beams in the paraxial approximation.
Without loss of generality, the beams propagate along $\pm{\rm z}$, respectively.
More specifically, we consider two linearly polarized laser beams of different peak field strength ${\cal E}_{0\pm}$, but of the same frequency $\Omega=\frac{2\pi}{\lambda}$ and pulse duration $\tau$, focused to the same beam waist $w_0$. 
The beam axes and foci of the two beams are assumed to be perfectly aligned, and we assume an optimal temporal overlap of the pulses in the focus.

\begin{figure}
 \centering
  \includegraphics[width=0.2\textwidth]{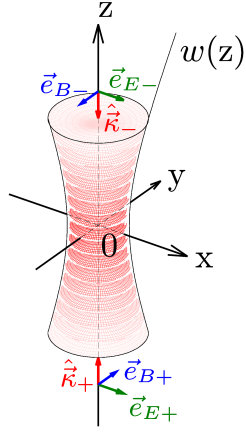} 
\caption{Schematic depiction of the background field configuration considered here. Two pulsed Gaussian laser beams propagating along $\hat{\vec{\kappa}}_\pm=\pm\vec{e}_{\rm z}$ are superposed in a counter-propagation geometry.
The associated electric and magnetic field vectors point in $\vec{e}_{E\pm}$ and $\vec{e}_{B\pm}$ directions; we depict the situation where $\phi_\pm=0$.
The widening of the beam's transverse extend as a function of $\rm z$ is described by $w({\rm z})=w_0\sqrt{1+(\frac{\rm z}{{\rm z}_R})^2}$; the beam waist is at ${\rm z}=0$.}
\label{fig:fig1}
\end{figure}

The electric $\vec{E}_\pm(x)={\cal E}_\pm(x)\vec{e}_{E\pm}$ and magnetic $\vec{B}_\pm(x)={\cal E}_\pm(x)\vec{e}_{B\pm}$ fields of the beams are orthogonal to the propagation direction $\pm\vec{e}_{\rm z}$ 
and are given by
$\vec{e}_{E\pm}=(\cos\phi_\pm,\pm\sin\phi_\pm,0)$ and $\vec{e}_{B\pm}=(-\sin\phi_\pm,\pm\cos\phi_\pm,0)$.
Each beam is characterized by a single amplitude profile, which can be represented as \cite{Siegman}
\begin{equation}
 {\cal E}_\pm(x)={\cal E}_{0\pm}\,{\rm e}^{-\frac{(\pm{\rm z}-t)^2}{(\tau/2)^2}}\frac{w_0}{w({\rm z})} {\rm e}^{-\frac{{\rm x}^2+{\rm y}^2}{w^2({\rm z})}}
\cos\bigl(\Phi_\pm(x)\bigr), \label{eq:env}
\end{equation}
with $\Phi_\pm(x)=\Omega(\pm{\rm z}-t)\pm\frac{\rm z}{{\rm z}_R}\frac{{\rm x}^2+{\rm y}^2}{w^2({\rm z})}\mp\arctan\bigl(\tfrac{\rm z}{{\rm z}_R}\bigr)+\varphi_{0\pm}$; see. Fig.~\ref{fig:fig1} for an illustration.
Here, $w({\rm z})=w_0\sqrt{1+(\frac{\rm z}{{\rm z}_R})^2}$ describes the widening of the beam's transverse extent as a function of $\rm z$, and $\arctan(\frac{\rm z}{{\rm z}_R})$ is the Gouy phase at $\rm z$.
Moreover, ${\rm z}_R=\frac{\pi w_0^2}{\lambda}$ denotes the Rayleigh range which is the distance from the focus (along $\rm z$) for which the beam radius is increased by a factor of $\sqrt{2}$;
$\varphi_{0\pm}$ is a constant phase.
An alternative representation of Eq.~\eqref{eq:env} is \cite{Karbstein:2015cpa}
\begin{equation}
 {\cal E}_\pm(x)=\frac{{\cal E}_{0\pm}}{2}\,{\rm e}^{-\frac{(\pm{\rm z}-t)^2}{(\tau/2)^2}} 
 \,\sum_{l=\pm1}\frac{1}{1\pm{\rm i}l\frac{{\rm z}}{{\rm z}_R}} \, {\rm e}^{-\frac{{\rm x}^2+{\rm y}^2}{w_0^2(1\pm{\rm i}l\frac{\rm z}{{\rm z}_R})}}{\rm e}^{{\rm i}l[\Omega(\pm{\rm z}-t)+\varphi_{0\pm}]}\,. \label{eq:Elin}
\end{equation}
For the considered background field configuration, the field invariants can be expressed as ${\cal F}=-2{\cal E}_+(x){\cal E}_-(x)\cos(\phi_++\phi_-)$ and ${\cal G}=2{\cal E}_+(x){\cal E}_-(x)\sin(\phi_++\phi_-)$.
It is moreover useful to introduce a normalized field strength tensor $\hat{F}^{\mu\nu}_\pm$ as follows ${\cal E}_\pm(x)\hat{F}^{\mu\nu}_\pm\equiv F^{\mu\nu}_\pm$.
For linearly polarized beams, this quantity does not vary as a function of space-time coordinates. It is fully characterized by the background field alignment with $e_{E\pm}^\mu\equiv(0,\vec{e}_{E\pm})$, $e_{B\pm}^\mu\equiv(0,\vec{e}_{B\pm})$,
$\hat\kappa_\pm^\mu=(1,\vec{e}_{E\pm}\times\vec{e}_{B\pm})=(1,\pm\vec{e}_{\rm z})$, and the in- and outgoing photon momenta $k'^\mu=(\omega',\vec{k}')$ and $k^\mu=(\omega,\vec{k})$.
In these notations we have
\begin{align}
 (k\hat F_\pm)^{\mu}&=(k\hat\kappa_\pm)e^\mu_{E\pm}-(k e_{E\pm})\hat\kappa_\pm^\mu\,, \nonumber\\
 (k{}^*\hat F_\pm)^{\mu}&=(k\hat\kappa_\pm)e^\mu_{B\pm}-(k e_{B\pm})\hat\kappa_\pm^\mu\,.
\end{align}
Finally, it is also useful to introduce the four momentum orthogonal to the propagation directions of the beams as $k_\perp^\mu=(0,k_{\rm x},k_{\rm y},0)$.
In turn, the photon polarization tensor in this background field configuration can be expressed in terms of the following analytical expression,
\begin{multline}
 \Pi^{\rho\sigma}(k,k')
 =\frac{\alpha}{\pi}\frac{1}{45}\Bigl(\frac{e}{m^2}\Bigr)^2
 \sum_{i=\pm}\bigl[4(k\hat F_i)^\rho  (k'\hat F_i)^\sigma + 7(k{}^*\hat F_i)^\rho (k'{}^*\hat F_i)^\sigma \bigr]{\cal I}_i(k,k')  \\
 -\frac{\alpha}{\pi}\frac{1}{45}\Bigl(\frac{e}{m^2}\Bigr)^2\Bigl\{
 8\cos(\phi_++\phi_-)\bigl((k k')g^{\rho\sigma} - k'^\rho k^\sigma \bigr)
 - 14\sin(\phi_++\phi_-)\epsilon^{\rho\sigma\alpha\beta} k'_\alpha k_\beta \\
 - 4\bigl[ (k\hat F_+)^\rho  (k'\hat F_-)^\sigma + (k\hat F_-)^\rho  (k'\hat F_+)^\sigma \bigr] \\
 - 7\bigl[ (k{}^*\hat F_+)^\rho (k'{}^*\hat F_-)^\sigma  + (k{}^*\hat F_-)^\rho (k'{}^*\hat F_+)^\sigma  \bigr]
 \Bigr\}{\cal J}(k,k') + {\cal O}(\epsilon^4)\,, \label{eq:Pi2}
\end{multline}
where the details of the field profile are encoded in the scalar functions
\begin{multline}
 {\cal I}_\pm(k,k')\equiv\int_x\,{\rm e}^{{\rm i}(k+k')x}\,{\cal E}_\pm^2(x)
 ={\cal E}_{0\pm}^2\,w_0^2\frac{\tau}{2}{\rm z}_R \Bigl(\frac{\pi}{2}\Bigr)^2 \\
 \times\biggl[\frac{2}{\sqrt{w_0^2(k_\perp+k_\perp')^2}}\,{\rm e}^{-2\frac{{\rm z}_R^2[(k+k')\hat\kappa_\pm]^2}{w_0^2(k_\perp+k_\perp')^2}-\frac{1}{8}[(\frac{\tau}{2})^2(\omega+\omega')^2+w_0^2(k_\perp+k_\perp')^2]} \\
 + \sqrt{\frac{\pi}{2}}\sum_{l=\pm1}{\rm e}^{-\frac{1}{8}(\frac{\tau}{2})^2(\omega+\omega'+2l\Omega)^2-l{\rm z}_R(k+k')\hat\kappa_\pm} {\rm e}^{{\rm i}l2\varphi_{0\pm}} \\
\times\Theta\Bigl(l{\rm z}_R(k+k')\hat\kappa_\pm-\tfrac{1}{8}w_0^2(k_\perp+k_\perp')^2\Bigr) \biggr] ,
\end{multline}
and
\begin{multline}
 {\cal J}(k,k')\equiv\int{\rm d}^4x\,{\rm e}^{{\rm i}(k+k')x}\,{\cal E}_+(x){\cal E}_-(x) ={\cal E}_{0+}{\cal E}_{0-} w_0^2\frac{\tau}{2}{\rm z}_R\Bigl(\frac{\pi}{2}\Bigr)^2 \\
 \times\sum_{l=\pm1}\Biggl\{
 \frac{1}{\sqrt{w_0^2(k_\perp+k_\perp')^2+16(\frac{2}{\tau}{\rm z}_R)^2}}\,{\rm e}^{-2\frac{{\rm z}^2_R(k_{\rm z}+k_{\rm z}')^2}{w_0^2(k_\perp+k_\perp')^2+16(\frac{2}{\tau}{\rm z}_R)^2}} \\
 \times {\rm e}^{-\frac{1}{8}[(\frac{\tau}{2})^2(\omega+\omega'+2l\Omega)^2+w_0^2(k_\perp+k'_\perp)^2]}{\rm e}^{{\rm i}l(\varphi_{0+}+\varphi_{0-})} \\
 +\sqrt{\frac{\pi}{2}}\,{\rm e}^{-\frac{1}{8}(\frac{\tau}{2})^2(\omega+\omega')^2+2(\frac{2}{\tau}{\rm z}_R)^2-{\rm z}_R[l(k_{\rm z}+k'_{\rm z})+2\Omega]}{\rm e}^{{\rm i}l(\varphi_{0+}-\varphi_{0-})} \\
 \times\frac{1}{2}\biggl[1+{\rm erf}\biggl(\frac{{\rm z}_R[2\Omega+l(k_{\rm z}+k_{\rm z}')]-\frac{1}{8}w_0^2(k_\perp+k'_\perp)^2}{2\sqrt{2}\,\frac{\tau}{2}{\rm z}_R}-\sqrt{2}\,\frac{\tau}{2}{\rm z}_R\biggr)\biggr]\Biggr\} .
\end{multline}
The expression in the first line of Eq.~\eqref{eq:Pi2} corresponds to the sum of the individual photon polarization tensors in pulsed Gaussian beams propagating in $\pm{\rm z}$ direction, respectively; cf. \cite{Karbstein:2015cpa}.
The additional terms are $\sim{\cal E}_{0+}{\cal E}_{0-}$ and hence describe effective couplings to the electromagnetic fields of both beams.
Note the much richer tensor structure for the superposition of two pulsed Gaussian beams as compared to that for a single Gaussian beam.
Of course, in the limit of ${\cal E}_{0\mp}\to 0$ the polarization tensor in a single Gaussian beam propagating along $\pm{\rm z}$ is recovered.

The expression derived here is, e.g., relevant for the study of optical signatures of quantum vacuum nonlinearity in the superposition of two counter-propagating high-intensity pump laser pulses.
It can, e.g., serve as an important input for a study of vacuum birefringence along the lines of \cite{Karbstein:2015xra,Karbstein:2016lby} in this alternative background field configuration,
and can facilitate a detailed study aiming at clarifying the question which field configuration maximizes the birefringence signal for a given pump laser energy:
By adjusting the field strengths of the individual beams accordingly, the background field configuration adopted here allows for a continuous deformation from the limit of a single propagating Gaussian beam, to the superposition of two counter-propagating Gaussian beams of equal strength which more closely resemble a standing-wave background field configuration.

\ack

I am grateful to the organizers of the International Workshop ``Strong Field Problems in Quantum Theory'' for organizing a very nice and stimulating conference at Tomsk State University.
Moreover, I am indebted to Elena~Mosman for exceptional excursions in Tomsk and to Holger~Gies for useful comments on this manuscript.



\section*{References}

\begin{thebibliography}{10}\setlength{\itemsep}{-0.5mm}

\bibitem{Heisenberg:1935qt} 
  W.~Heisenberg and H.~Euler,
  Z.\ Phys.\  {\bf 98}, 714 (1936), 
  an English translation is available at [physics/0605038].

\bibitem{Cangemi:1995ee} 
  D.~Cangemi, E.~D'Hoker and G.~V.~Dunne,
  Phys.\ Rev.\ D {\bf 52}, 3163 (1995)
  [hep-th/9506085].
  
\bibitem{Dunne:1997kw} 
  G.~V.~Dunne and T.~M.~Hall,
  Phys.\ Lett.\ B {\bf 419}, 322 (1998)
  [hep-th/9710062].

\bibitem{Dunne:1998ni} 
  G.~V.~Dunne and T.~Hall,
  Phys.\ Rev.\ D {\bf 58}, 105022 (1998)
  [hep-th/9807031].
  
\bibitem{Dunne:2004nc} 
  G.~V.~Dunne,
  In *Shifman, M. (ed.) et al.: From fields to strings, vol. 1* 445-522
  [hep-th/0406216].
  
\bibitem{Schwinger:1951nm} 
  J.~S.~Schwinger,
  Phys.\ Rev.\  {\bf 82}, 664 (1951).
  
\bibitem{BatShab}
  I.~A.~Batalin and A.~E.~Shabad,
  Zh.\ Eksp.\ Teor.\ Fiz.\  {\bf 60}, 894 (1971)
  [Sov.\ Phys.\ JETP\ {\bf 33}, 483 (1971)].
  
\bibitem{Baier:1975ff} 
  V.~N.~Baier, A.~I.~Milshtein and V.~M.~Strakhovenko,
  Zh.\ Eksp.\ Teor.\ Fiz.\  {\bf 69}, 1893 (1975)
  [Sov.\ Phys.\ JETP\ {\bf 42}, 961 (1976)].
  
\bibitem{Becker:1974en} 
  W.~Becker and H.~Mitter,
   J.\ Phys.\ A:\ Math.\ Gen.\ {\bf 8} 1638 (1975).

\bibitem{DiPiazza:2011tq} 
  A.~Di Piazza, C.~Muller, K.~Z.~Hatsagortsyan and C.~H.~Keitel,
  Rev.\ Mod.\ Phys.\  {\bf 84}, 1177 (2012)
  [arXiv:1111.3886 [hep-ph]].
  
\bibitem{Siegman}
A.~E.~Siegman, \textit{Lasers}, First Edition, University Science Books, USA (1986);
B.~E.~A.~Saleh and M.~C.~Teich, \textit{Fundamentals of Photonics}, First Edition, John Wiley \& Sons, USA (1991). 
  
\bibitem{Karbstein:2015cpa} 
  F.~Karbstein and R.~Shaisultanov,
  Phys.\ Rev.\ D {\bf 91}, no. 8, 085027 (2015)
  [arXiv:1503.00532 [hep-ph]].
  
\bibitem{Dittrich:1985yb} 
  W.~Dittrich and M.~Reuter,
  Lect.\ Notes Phys.\  {\bf 220}, 1 (1985);
  W.~Dittrich and H.~Gies,
  Springer Tracts Mod.\ Phys.\  {\bf 166}, 1 (2000).

\bibitem{Karbstein:2015xra} 
  F.~Karbstein, H.~Gies, M.~Reuter and M.~Zepf,
  Phys.\ Rev.\ D {\bf 92}, 071301 (2015)
  [arXiv:1507.01084 [hep-ph]].

\bibitem{Karbstein:2016lby} 
  F.~Karbstein and C.~Sundqvist,
  arXiv:1605.09294 [hep-ph].
  
\end{thebibliography}
\end{document}